\documentclass[
 reprint,
 superscriptaddress,
 amsmath,amssymb,
 aps,
]{revtex4-2}

\usepackage{graphicx}
\usepackage{dcolumn}
\usepackage{bm}
\usepackage{color}

\begin{document}

\preprint{APS/123-QED}

\title{Diffusion enhancement in a levitated droplet via oscillatory deformation}

\author{Yuki Koyano}
\email{koyano@cmpt.phys.tohoku.ac.jp}
\affiliation{Department of Physics, Graduate School of Science, Tohoku University, Sendai, Miyagi 980-8578, Japan}

\author{Hiroyuki Kitahata}
\email{kitahata@chiba-u.jp}
\affiliation{Department of Physics, Graduate School of Science, Chiba University, Chiba 263-8522, Japan}

\author{Koji Hasegawa}
\affiliation{Faculty of Engineering, Kogakuin University, Shinjuku-ku, Tokyo 163-8677, Japan}

\author{Satoshi Matsumoto}
\affiliation{Human Space Flight Technology Directorate, Japan Space Exploration Agency, Tsukuba, Ibaraki 305-8505, Japan}

\author{Katsuhiro Nishinari}
\affiliation{Research Center for Advanced Science and Technology, The University of Tokyo, Meguro-ku, Tokyo 153-8904, Japan}

\author{Tadashi Watanabe}
\affiliation{Research Institute of Nuclear Engineering, University of Fukui, Tsuruga, Fukui 914-0055, Japan}

\author{Akiko Kaneko}
\affiliation{Graduate School of System and Information Engineering, University of Tsukuba, Tsukuba, Ibaraki 305-8573, Japan}

\author{Yutaka Abe}
\affiliation{Graduate School of System and Information Engineering, University of Tsukuba, Tsukuba, Ibaraki 305-8573, Japan}

\date{\today}

\begin{abstract}
Recent experimental results indicate that mixing is enhanced by a reciprocal flow induced inside a levitated droplet with an oscillatory deformation [T.~Watanabe \textit{et al.} Sci. Rep. \textbf{8}, 10221 (2018)].
Generally, reciprocal flow cannot convect the solutes in time average, and agitation cannot take place.
In the present paper, we focus on the diffusion process coupled with the reciprocal flow.
We theoretically derive that the diffusion process can be enhanced by the reciprocal flow, and the results are confirmed via numerical calculation of the over-damped Langevin equation with a reciprocal flow.
\end{abstract}

\maketitle

\section{Introduction}

A levitated-droplet system was intensively developed to realize contactless manipulation.
For instance, it was advantageous to measure physical or chemical quantities by avoiding the pollution and the significant disturbance from chamber walls~\cite{Ishikawa}.
Thus, levitated-droplet systems were studied not only under microgravity conditions in a spacecraft or during a parabolic flight but also in various conditions, i.e., electrostatic levitation~\cite{Rhim, Brillo}, magnetic levitation~\cite{Liua, Hill, Mestel}, acoustic levitation~\cite{Marzo, Hirayama, Morris, Sasaki, OhsakaPRL}, aerodynamic levitation~\cite{Feng}, and optical levitation~\cite{Price}.
For each levitation method, the droplet system had to be designed for the stabilization of the levitated state.
Furthermore, in relation to the levitation techniques, the dynamics of a droplet were also studied intensively~\cite{DynamicsDroplet, Watanabe2008, Watanabe2010, KitahataPRE, Mestel}.

In this study, we focus on acoustic levitation, where a droplet can be levitated at the valleys of a standing wave in a three-dimensional sound pressure field~\cite{Marzo, Hirayama, Morris}.
By using the arrays of the sound sources and controlling the phases of the irradiated sound, multiple droplets can be levitated simultaneously and their positions can be controlled.
This enables merging of two or more levitated droplets with contactless operations, which will be of significant importance for syntheses of materials without any contact with the apparatuses.

Recently, a contactless mixing technique in an acoustic levitation system was reported~\cite{Watanabe, Hasegawa}.
By applying the frequency modulation to the sound pressure field, the levitated droplet can exhibit an oscillatory deformation.
By observing the time evolution of the fluorescent dye distribution, it was concluded that such deformation enhanced the mixing inside the droplet.
The flow inside the droplet is also observed via particle image velocimetry (PIV).
The observed flow field appeared reciprocal. 
The reciprocal flow cannot convect the solutes in time average; thus, agitation cannot occur.
This is known as the scallop theorem~\cite{Purcell}.
Thus, the mixing enhancement should not originate from the agitation by the flow. However, it should originate from the diffusion enhancement by the reciprocal flow.
Since the diffusion enhancement by reciprocal flow was found heuristically, the clarification of the mechanism is awaited.
The mechanism may contribute to the development of the mixing technique for designing more efficient systems.

The diffusion process affected by the reciprocal flow field is a nontrivial dynamics.
To date, the dynamics of agitation by flow and that of diffusion originating from thermal fluctuation were separately considered. 
This is because they have different time scales and spatial scales.
The mixing by agitation proceeds faster on a larger spatial scale, while the diffusion proceeds faster on a smaller spatial scale.
Furthermore, in the convection-diffusion equation, the terms describing the convection and diffusion are separately described.
It should be noted that the reciprocal flow does not cause the Stokes drift~\cite{Stokes1847}; thus, we do not consider it.

In the present paper, we study the effect of the reciprocal flow field on the diffusion process.
In Sec.~II, we construct a solution for the reciprocal flow field.
Subsequently, we formulate a diffusion equation per every period of the oscillatory deformation in Sec.~III.
We found that the diffusion coefficient includes not only the classical diffusion coefficient but also a combination term of diffusion and flow field.
The latter term indicates the diffusion enhancement.
The theoretical result is confirmed via numerical calculations in Sec.~IV.
In Sec.~V, the validity of the adopted assumptions is verified.
Finally, we discuss the physical meaning of the diffusion coupled with the reciprocal flow as the summary.

\section{Flow field in a droplet}

First, the model equations for the droplet with an oscillatory deformation are introduced.
The flow in the droplet, $\bm{v}(\bm{r},t)$, is described by the Navier-Stokes equation~\cite{Lamb,Landau}:
\begin{equation}
\varrho \left(\frac{\partial \bm{v}}{\partial t} + (\bm{v} \cdot \nabla) \bm{v} \right) = -\nabla p + \eta \nabla^2 \bm{v} \label{NS}
\end{equation}
with incompressibility:
\begin{equation}
\nabla \cdot \bm{v} = 0, \label{incompressibility}
\end{equation}
where $\varrho$ and $\eta$ denote the density and viscosity of the fluid, respectively.
Furthermore, $p(\bm{r}, t)$ denotes the pressure inside the droplet.

To describe the droplet, the following two boundary conditions are imposed. The first condition is with respect to the balance of pressure at the droplet surface as follows:
\begin{equation}
\left. p\right|_{{\bm r} = \bm{r}_b} = p_{\mathrm{air}} - 2 \gamma \left. H \right|_{{\bm r} = \bm{r}_b} + p_{\mathrm{acoustic}}, \label{LaplacePressure}
\end{equation}
where $p_\mathrm{air} (= \mathrm{const.})$ denotes the atmospheric pressure, $\gamma$ denotes the surface tension, and $H$ denotes the mean curvature of the droplet surface.
The explicit description of $H$ is denoted in Appendix~\ref{mc}.
The vector $\bm{r}_b = f(\theta, \varphi, t) \bm{e}_r$ represents a point on the droplet surface in polar coordinates $(r, \theta, \varphi)$.
Here, $\bm{e}_r$ denotes the unit vector in the $r$ direction.
The second term in the right side of Eq.~\eqref{LaplacePressure} denotes the Laplace pressure, which originates from the minimization of the surface energy~\cite{deGennes}.
The deformation of the droplet affects the flow through this term.

The second boundary condition is for the flow field.
It states that the normal component of the flow $\bm{v}$ should be equal to that of the velocity of the droplet boundary as follows:
\begin{align}
\left. \bm{v}\right|_{{\bm r} = \bm{r}_b} \cdot \left. \bm{n}\right|_{{\bm r} = \bm{r}_b} = \frac{d \bm{r}_b}{dt} \cdot \left.\bm{n}\right|_{{\bm r} = \bm{r}_b}, \label{bc_shape} 
\end{align}
where $\bm{n}$ denotes the outward normal unit vector at the droplet surface.

The variables are non-dimensionalized as follows: $\tilde{\bm{r}} = \bm{r}/R$, $\tilde{\bm{r}}_b = \bm{r}_b /R$, $\tilde{\bm{v}} = (T/R) \bm{v}$, $\tilde{t} = t/T$, $\tilde{\nabla} = R \nabla$, $\tilde{p} = p / p_0 = p T^2 / (\varrho R^2)$, $\tilde{H} = R H$, and $\tilde{p}_{\mathrm{acoustic}} = p_{\mathrm{acoustic}} / p_0$, 
where $R$ denotes the radius of the droplet, and $T$ denotes the period of the oscillatory deformation.
The tilde signs $(\tilde{})$ indicate the dimensionless variables.
The dimensionless forms of equations are described as follows:
\begin{equation}
\frac{\partial \tilde{\bm{v}}}{\partial \tilde{t}} + (\tilde{\bm{v}} \cdot \tilde{\nabla}) \tilde{\bm{v}} = - \tilde{\nabla} \tilde{p} + \frac{1}{\mathrm{Re}} \tilde{\nabla}^2 \tilde{\bm{v}}, \label{NS_nd}
\end{equation}
\begin{align}
\left. \tilde{p}\right|_{\tilde{\bm{r}} = \tilde{\bm{r}}_b} = P_\mathrm{air} - 2 \sigma \left. \tilde{H} \right|_{\tilde{\bm{r}} = \tilde{\bm{r}}_b} + \tilde{p}_{\mathrm{acoustic}}, \label{LaplacePressure_nd}
\end{align}
where $\sigma = \gamma T^2/(\varrho R^3)$, $\mathrm{Re} = \varrho R^2 /(\eta T)$, and $P_\mathrm{air} = p_\mathrm{air} / p_0$ denote the dimensionless parameters.
Since the order of $1/\mathrm{Re}$ is $0.01$ in the experiments \cite{Watanabe}, we neglect the viscous term (momentum diffusion term).

In the experiments \cite{Watanabe}, the droplet periodically changes its shape keeping the oscillation amplitude.
The energy injection term $\tilde{p}_{\mathrm{acoustic}}$ should be of the same order as 0.01 because energy dissipation and injection should be balanced in time average.
Thus, $\tilde{p}_{\mathrm{acoustic}}$ should be also neglected because the viscous term $(1/\mathrm{Re}) \tilde{\nabla}^2 \tilde{\bm{v}}$ is neglected.
Hereafter, we treat the following equations:
\begin{equation}
\frac{\partial \tilde{\bm{v}}}{\partial \tilde{t}} + (\tilde{\bm{v}} \cdot \tilde{\nabla}) \tilde{\bm{v}} = - \tilde{\nabla} \tilde{p}, \label{NS_mod-}
\end{equation}
\begin{equation}
\tilde{\nabla} \cdot \tilde{\bm{v}} = \bm{0}, \label{incompressibility_nd}
\end{equation}
\begin{align}
\left. \tilde{p}\right|_{\tilde{\bm{r}} = \tilde{\bm{r}}_b} = P_{\mathrm{air}} - 2\sigma \left. \tilde{H} \right|_{\tilde{\bm{r}} = \tilde{\bm{r}}_b}, \label{LaplacePressure_mod}
\end{align}
\begin{align}
\left. \tilde{\bm{v}} \right|_{\tilde{\bm{r}} = \tilde{\bm{r}}_b} \cdot \left. \bm{n}\right|_{\tilde{\bm{r}} = \tilde{\bm{r}}_b} = \frac{d \tilde{\bm{r}}_b}{d\tilde{t}} \cdot \left.\bm{n}\right|_{\tilde{\bm{r}} = \tilde{\bm{r}}_b}, \label{bc_shape-} 
\end{align}
For simplicity, we omit the tildes in the following descriptions.

By assuming that the flow has no vorticity, the flow field can be represented by the velocity potential $\Phi(\bm{r},t)$ as
\begin{align}
\bm{v} = \nabla \Phi. \label{def_phi}
\end{align}
Subsequently, Eq.~\eqref{NS_mod-} is represented as follows:
\begin{align}
\frac{\partial \Phi}{\partial t} + \frac{1}{2} \left| \nabla \Phi \right|^2 = -p. \label{NS_mod}
\end{align}
Equations~\eqref{incompressibility_nd} and \eqref{def_phi} lead
\begin{align}
\nabla^2 \Phi = 0, \label{harmonic}
\end{align}
which indicates that $\Phi$ is a harmonic function.

In the experiments, the amplitude of the shape oscillation $\Delta R$ is small, which is characterized by the dimensionless parameter $\varepsilon = \Delta R / R$.
Here, we used the perturbation method; the solution is expanded with respect to a small parameter $\varepsilon$ up to the second order as follows:
\begin{align}
f &= f^{(0)} + \varepsilon f^{(1)} + \varepsilon^2 f^{(2)} + \mathcal{O}( \varepsilon^3), \label{f_expand} \\
\Phi &= \Phi^{(0)} + \varepsilon \Phi^{(1)} + \varepsilon^2 \Phi^{(2)} + \mathcal{O}( \varepsilon^3), \label{Phi_expand} \\
p &= p^{(0)} +\varepsilon p^{(1)} +\varepsilon^2 p^{(2)} + \mathcal{O}( \varepsilon^3). \label{p_expand}
\end{align}
By substituting the aforementioned expressions, we obtain the order-separated equations with respect to $\varepsilon$.
The equations are shown in Appendix~\ref{eq_bc}.
It is noted that the pressure $p$ can be easily eliminated from the equation and boundary conditions, and thus the deformation $f$ and the velocity potential $\Phi$ are calculated.

A trivial solution for the equation is as follows:
\begin{align}
\bm{r}_b &= f^{(0)} \bm{e}_r = \bm{e}_r, \\
\Phi &= \Phi^{(0)} = \Phi_0,
\end{align}
This corresponds to the solution for the static state of the droplet.
The corresponding pressure and flow field are described as follows:
\begin{align}
p &= p^{(0)} = P_\mathrm{air} + 2\sigma, \\
\bm{v} &= \bm{v}_0 = \bm{0}.
\end{align}

When the droplet is deformed from the sphere, the droplet tends to return to the sphere because of the surface tension.
Since the surface tension works as a restoring force, the droplet exhibits a harmonic oscillation in the order of $\varepsilon$.
The generic solution $f_\mathrm{gen}^{(1)}$ and $\Phi_\mathrm{gen}^{(1)}$ for the flow and deformation is explicitly described using the spherical harmonics as follows:
\begin{align}
f^{(1)}_\mathrm{gen} 
=& \left[ \sum_{\ell=2}^\infty \sum_{m = 0}^\ell \beta^{(1c)}_{\ell, m} P_\ell^{\left| m \right|}(\cos\theta) \cos m \varphi \sin(\omega_\ell t + \delta_\ell) \right. \nonumber \\
&\left. + \sum_{\ell=2}^\infty \sum_{m = 1}^\ell \beta^{(1s)}_{\ell, m} P_\ell^{\left| m \right|}(\cos\theta) \sin m \varphi \sin(\omega_\ell t + \delta_\ell) \right], \label{r1_gen}
\end{align}
\begin{align}
&\Phi^{(1)}_\mathrm{gen} \nonumber \\
&= \left [ \sum_{\ell=2}^\infty \sum_{m = 0}^\ell \frac{\beta^{(1c)}_{\ell, m} \omega_\ell}{\ell} r^\ell P_\ell^{\left| m \right|}(\cos\theta) \cos m \varphi \cos(\omega_\ell t + \delta_\ell) \right . \nonumber \\
&\left . + \sum_{\ell=2}^\infty \sum_{m = 1}^\ell \frac{\beta^{(1s)}_{\ell, m} \omega_\ell}{\ell} r^\ell P_\ell^{\left| m \right|}(\cos\theta) \sin m \varphi \cos(\omega_\ell t + \delta_\ell) \right ], \label{Phi1_gen}
\end{align}
where $P_\ell^m$ is an associated Legendre polynomial of the degree $\ell$ and order $m$.
The amplitudes of the oscillation modes $\beta^{(1c)}_{\ell,m}$ and $\beta^{(1s)}_{\ell,m}$ are constants that are chosen arbitrarily.
Here, the characteristic frequency $\omega_\ell$ is described as follows:
\begin{align}
\omega_\ell = \sqrt{\sigma \ell (\ell - 1)(\ell + 2)}. \label{nat_freq}
\end{align}
The result in the order of $\varepsilon$ is consistent with the results by Rayleigh~\cite{Rayleigh}.
It is important to note that $\omega_\ell = 0$ for the modes of $\ell = 0$ and $1$.
The mode $\ell = 0$ corresponds to extension and contraction. We consider incompressible fluid; thus, the mode $\ell=0$ is not a physical solution.
The mode $\ell = 1$ does not correspond to a deformation. However, it corresponds to the oscillation or translation of the center position.
The surface tension can work as a restoring force for the deformation, but it cannot drift the droplet if $\gamma = \mathrm{const.}$ 
Additionally, we do not consider the case that the acoustic pressure field drifts the droplet.
Thus, the mode $\ell = 1$ should not appear.
Therefore, we only need to consider the modes for $\ell \ge 2$.

In the experiments \cite{Watanabe}, the droplet was observed both from the top and side.
The droplet deformed in the horizontal direction, but it did not deform in the vertical direction.
The flow at the equatorial plane was also observed via PIV.
Based on the observation of the shape change and PIV, the flow field appeared to correspond to a single mode $\ell = m = n$ $(n \in \mathbb{N}, n \ge 2)$.
Thus, hereafter, we adopted a single mode of $\ell = m = n$.
Moreover, only the cosine mode is considered for simplicity.
As for the time evolution, we can arbitrarily choose $\delta_n = 0$ owing to the time translational symmetry.
Based on the aforementioned assumptions, we have
\begin{align}
f^{(1)} =& \sin^n \theta \cos n \varphi \sin \omega_n t, \label{r1} \\
\Phi^{(1)} =& \frac{\omega_n}{n} r^n \sin^n \theta \cos n \varphi \cos \omega_n t. \label{Phi1}
\end{align}
Here we use
\begin{align}
P_n^n(\cos\theta) = \left[(-1)^n (2n-1)!! \right] \sin^n \theta,
\end{align}
and set
\begin{align}
\beta^{(1c)}_{n,n} = \frac{1}{(-1)^n (2n-1)!!},
\end{align}
such that the small parameter $\varepsilon$ denotes the amplitude of the oscillatory deformation.

Subsequently, we consider the flow field in the order of $\varepsilon^2$ in the case that the flow field in the order of $\varepsilon$ is expressed in Eqs.~\eqref{r1} and \eqref{Phi1}.
Based on the equations, Eqs.~\eqref{NS2}, \eqref{bc2}, and \eqref{phi2} in Appendix~\ref{eq_bc}, for the order of $\varepsilon^2$, we obtain
\begin{align}
&f^{(2)} = \hat{f}^{(2,n)}_{2n,2n} \sin^{2n}\theta \cos 2n\varphi \cos 2\omega_n t \nonumber \\
& + \bar{f}^{(2,n)}_{2n,2n} \sin^{2n}\theta \cos 2n\varphi + \sum_{k=0}^n \hat{f}^{(2,n)}_{2k,0} P_{2k}(\cos \theta) \cos 2\omega_n t \nonumber \\
& + \sum_{k=0}^n \bar{f}^{(2,n)}_{2k,0} P_{2k}(\cos \theta), \label{r2exp}
\end{align}
\begin{align}
\Phi^{(2)} =& \hat{\Phi}^{(2,n)}_{2n,2n} \left( r \sin \theta \right)^{2n} \cos 2n\varphi \sin 2 \omega_n t \nonumber \\
&+ \sum_{k=0}^{n} \hat{\Phi}^{(2,n)}_{2k,0} r^{2k} P_{2k}(\cos\theta) \sin 2 \omega_n t. \label{Phi2exp}
\end{align}
Here, we introduce the constants $\hat{f}^{(2,n)}_{2n,2n}$, $\bar{f}^{(2,n)}_{2n,2n}$, $\hat{f}^{(2,n)}_{2k,0}$, $\bar{f}^{(2,n)}_{2k,0}$, $\hat{\Phi}^{(2,n)}_{2n,2n}$, and $\hat{\Phi}^{(2,n)}_{2k,0}$, whose explicit forms are shown in Appendix~\ref{sol}.

\section{Diffusion with the reciprocal flow}

To describe the dynamics of the tracer particle inside an oscillatory deformed droplet, the over-damped Langevin equation, which is affected by the thermal noise and advection due to the reciprocal flow, is adopted as follows:
\begin{align}
\frac{d\bm{x}}{dt} = \bm{v}(\bm{x},t) + \bm{\xi}(t), \label{sd_eq}
\end{align}
where $\bm{x}$ denotes the position of the tracer particle and $\bm{v}(\bm{x},t)$ denotes the flow field obtained in the last section.
The flow field can be expressed as follows:
\begin{align}
\bm{v} =&\varepsilon \nabla \Psi^{(1)} \cos \omega_n t + \varepsilon^2 \nabla \Psi^{(2)} \sin 2\omega_n t + \mathcal{O}(\varepsilon^3).
\end{align}
Here, we define $\Psi^{(1)}$ and $\Psi^{(2)}$ such that
\begin{align}
\Phi^{(1)} (\bm{x},t) &= \Psi^{(1)} (\bm{x}) \cos \omega_n t \\
\Phi^{(2)} (\bm{x},t) &= \Psi^{(2)} (\bm{x}) \sin 2\omega_n t,
\end{align}
where $\Phi^{(1)}$ and $\Phi^{(2)}$ are explicitly provided in Eqs.~\eqref{Phi1} and \eqref{Phi2exp}, respectively.
The function $\bm{\xi}(t)$ corresponds to the thermal noise and satisfies the following relations:
\begin{align}
&\left < \xi_\alpha (t) \right > = 0, \\
&\left < \xi_\alpha (t) \xi_\beta (s)\right > = 2 D \delta_{\alpha \beta} \delta(t-s),
\end{align}
where $\delta_{\alpha \beta}$ denotes the Kronecker delta and $\delta(\cdot)$ denotes the Dirac's delta function, and $D$ is the diffusion coefficient originating from the thermal noise.

The Fokker-Planck equation for Eq.~\eqref{sd_eq} is derived as follows:
\begin{align}
\frac{\partial q (\bm{r},t)}{\partial t} = - \nabla \cdot \left ( \bm{v} (\bm{r},t) q (\bm{r},t) \right ) + D \nabla^2 q (\bm{r},t), \label{FP}
\end{align}
where $q (\bm{r},t)$ denotes the probability density of the tracer particle~\cite{Risken, Synergetics2}.
In this Fokker-Planck equation, the effects of the convection and the diffusion appear separately.
Here, we consider the map of the probability density $q (\bm{r},t)$ per the period of the flow field, instead of the Fokker-Planck equation.
The map for the probability density $\hat{q}_{j} (\bm{r}) = q (\bm{r}, j)$ ($\hat{q}_{j} (\bm{r}) = q (\bm{r}, jT)$ with the time dimension) is described as follows:
\begin{align}
&\hat{q}_{j+1} (\bm{r}) - \hat{q}_{j} (\bm{r}) \nonumber \\
&= - \frac{\partial}{\partial x_\alpha} \left ( M_\alpha^{(1)} (\bm{r}) \hat{q}_{j} (\bm{r}) \right ) + \frac{1}{2} \frac{\partial^2}{\partial x_\alpha \partial x_\beta} \left ( M_{\alpha \beta}^{(2)} (\bm{r}) \hat{q}_{j} (\bm{r}) \right ), \label{dFP}
\end{align}
where $j \in \mathbb{N}$ indicates the index of the period.
Equation \eqref{dFP} is considered as a discrete Fokker-Planck equation.
The higher-order spatial derivatives of $\hat{q}$ are neglected.
It should be noted that Eq.~\eqref{dFP} is valid when the time evolution per the period, $\hat{q}_{j+1} (\bm{r}) - \hat{q}_{j} (\bm{r})$, is sufficiently small.
The first and second moments $M_\alpha^{(1)} (\bm{r})$ and $M_{\alpha\beta}^{(2)} (\bm{r})$ in Eq.~\eqref{dFP}, respectively, are defined as the alternatives of the Kramers-Moyal coefficients as follows:
\begin{align}
M_\alpha^{(1)}(\bm{r}) =& \left < \Delta x_\alpha \right >, \label{def_KM1} \\
M_{\alpha \beta}^{(2)}(\bm{r}) =& \left < \Delta x_\alpha \Delta x_\beta \right >, \label{def_KM2} 
\end{align}
where $\Delta x_\alpha$ denotes the displacement in the period of the oscillation.
They are calculated as follows:
\begin{align}
M_\alpha^{(1)} (\bm{r}) = \frac{D \varepsilon^2}{2{\omega_n}^2} \frac{\partial^3 \Psi^{(1)} (\bm{r})}{\partial x_\alpha \partial x_{\alpha'} \partial x_{\alpha''}} \frac{\partial^2 \Psi^{(1)} (\bm{r})}{\partial x_{\alpha'} \partial x_{\alpha''}} + \mathcal{O}(\varepsilon^3, D^2), \label{KM1}
\end{align}
\begin{widetext}
\begin{align}
M_{\alpha \beta}^{(2)} (\bm{r}) 
=& 2D \left [ \delta_{\alpha \beta} + \frac{\varepsilon^2}{2{\omega_n}^2} \frac{\partial}{\partial x_{\alpha'}} \left ( \frac{\partial \Psi^{(1)} (\bm{r})}{\partial x_\alpha} \frac{\partial^2 \Psi^{(1)} (\bm{r})}{\partial x_{\alpha'} \partial x_{\beta}} 
+ \frac{\partial \Psi^{(1)} (\bm{r})}{\partial x_\beta} \frac{\partial^2 \Psi^{(1)} (\bm{r})}{\partial x_{\alpha'} \partial x_{\alpha}} 
- \frac{\partial \Psi^{(1)} (\bm{r})}{\partial x_{\alpha'}} \frac{\partial^2 \Psi^{(1)} (\bm{r})}{\partial x_\alpha \partial x_\beta} \right ) - \frac{\varepsilon^2}{\omega_n} \frac{\partial^2 \Psi^{(2)} (\bm{r})}{\partial x_\alpha \partial x_\beta} \right ] \nonumber \\
& + \mathcal{O}(\varepsilon^3, D^2). \label{KM2}
\end{align}
\end{widetext}
Here, the terms with the order of $\varepsilon^3$ and $D^2$ or higher are neglected.
By substituting Eqs.~\eqref{KM1} and \eqref{KM2} into Eq.~\eqref{dFP}, we have
\begin{align}
&\hat{q}_{j+1} (\bm{r}) - \hat{q}_{j} (\bm{r}) = \frac{\partial}{\partial x_\alpha} \left ( D_{\alpha \beta}^{\mathrm{eff}} (\bm{r}) \frac{\partial \hat{q}_{j} (\bm{r})}{\partial x_\beta} \right ), \label{dfp.eq}
\end{align}
where $D_{\alpha \beta}^{\mathrm{eff}} (\bm{r})$ is defined as
\begin{align}
D_{\alpha \beta}^{\mathrm{eff}} (\bm{r}) = \frac{1}{2} M_{\alpha \beta}^{(2)} (\bm{r}). \label{D_M}
\end{align}
Equation~\eqref{dfp.eq} shows that only the diffusion term remains.
It is important to note that the diffusion tensor $D^{\mathrm{eff}}$ is not only dependent on the thermal diffusion coefficient $D$ but also on the flow field.
Thus, it exhibits the spatial dependence $D^{\mathrm{eff}} = D^{\mathrm{eff}}(\bm{r})$.

Here we consider the mass diffusion on the $xy$-plane at $z=0$, where the deformation is the largest.
The diffusion tensor in the cylindrical coordinates $(\rho, \varphi, z)$ is considered by reflecting the system symmetry.
The diffusion tensor is defined in Eqs.~\eqref{def_drr} to \eqref{def_dphiz} in Appendix~\ref{exp_D}.
The effective diffusion coefficient $\bar{D}^{\mathrm{eff}}$ is defined as follows:
\begin{align}
\bar{D}^{\mathrm{eff}} = \frac{1}{3} \left(D^{\mathrm{eff}}_{\rho\rho} + D^{\mathrm{eff}}_{\varphi \varphi} + D^{\mathrm{eff}}_{zz}\right). \label{eff_diff}
\end{align}
Based on the theoretical calculation, it is obtained as follows:
\begin{align}
\frac{\bar{D}^{\mathrm{eff}}}{D} = 1 + \varepsilon^2 a_n \rho^{2(n-2)}, \label{D_eff}
\end{align}
where
\begin{align}
a_n = \frac{2}{3} (n-1)^2.
\end{align}
Since $a_n$ is positive for $n \geq 2$, the diffusion is enhanced for all the modes of the deformation.
In the case of $n=2$, the effective diffusion coefficient $\bar{D}^{\mathrm{eff}}$ does not have a spatial dependence, reflecting that the averaged shear strain is uniform at any point in the droplet~\cite{KitahataPRE}.
Conversely, in the case of $n \geq 3$, the effective diffusion coefficient $\bar{D}^{\mathrm{eff}}$ is larger near the surface of the droplet.
This result is reasonable because the deformation of the fluid element is large near the surface of the droplet.

Below, each component in the diffusion tensor is given as follows:
\begin{align}
\frac{D^{\mathrm{eff}}_{\rho\rho}}{D} =& 1 + \varepsilon^2 \left[ b_n + \sum_{k=2}^{n} c_{nk} \rho^{2(k-1)} + d_n \rho^{2(n-1)}\cos (2n\varphi) \right], \label{drr}
\end{align}
\begin{align}
\frac{D^{\mathrm{eff}}_{\varphi \varphi}}{D} =& 1 + \varepsilon^2 \left[ b_n + \sum_{k=2}^{n} g_{nk} \rho^{2(k-1)} - d_n \rho^{2(n-1)} \cos (2n\varphi) \right], \label{dphiphi}
\end{align}
\begin{align}
\frac{D^{\mathrm{eff}}_{zz}}{D} = 1 + \varepsilon^2 \sum_{k=1}^{n} h_{nk} \rho^{2(k-1)}, \label{dzz}
\end{align}
\begin{align}
\frac{D^{\mathrm{eff}}_{\rho\varphi}}{D} = - \varepsilon^2 d_n \rho^{2(n-1)} \sin (2n\varphi). \label{drphi}
\end{align}
The other components $D^{\mathrm{eff}}_{\rho z}$ and $D^{\mathrm{eff}}_{\varphi z}$ are zero.
Here, $b_n$, $c_{nk}$, $d_{n}$, $g_{nk}$, and $h_{nk}$ are constants, which are explicitly described in Appendix~\ref{exp_D}.
The effective diffusion tensor $D^{\mathrm{eff}}$ is not diagonal.
This implies that the diffusion is anisotropic.

\section{Numerical simulation}

\begin{figure}
	\centering
	\includegraphics{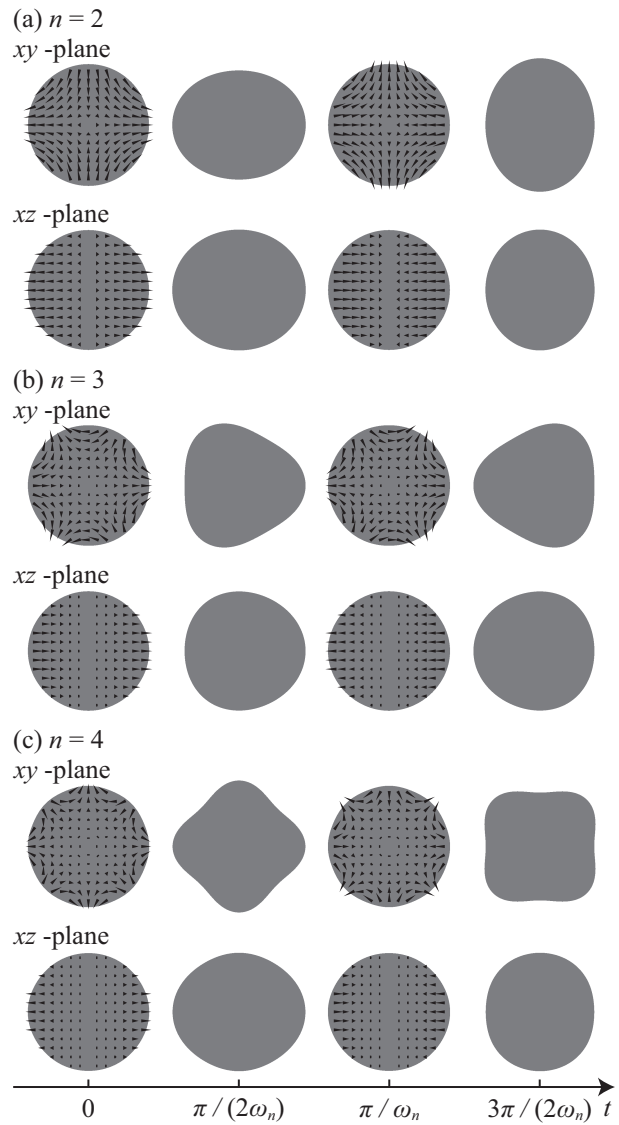}
	\caption{Time series of the droplet shape and flow field. The gray regions illustrate the cross section of the droplet at $xy$- and $xz$- planes. The black arrowheads represent the flow fields.}
	\label{fig_snapshots}
\end{figure}

The numerical calculations were performed to confirm the theoretical results. 
The over-damped Langevin equation in Eq.~\eqref{sd_eq} for a tracer particle was adopted.
In the calculation, we set the parameters to be $\omega_n = 1$ and $D = 10^{-5}$. The calculation was executed with the second-order Runge-Kutta method (explicit midpoint method), and the Gaussian white noise was generated with the Box-Muller method~\cite{Numerical_Recipes}.
The time step was set as $\Delta t = 1 / 400$.
With respect to the flow field, $\bm{v}$, we used the theoretically obtained flow field.
The examples of the flow field for mode $n = 2, 3$, and $4$ are shown in Fig.~\ref{fig_snapshots}.

\begin{figure}
	\centering
	\includegraphics{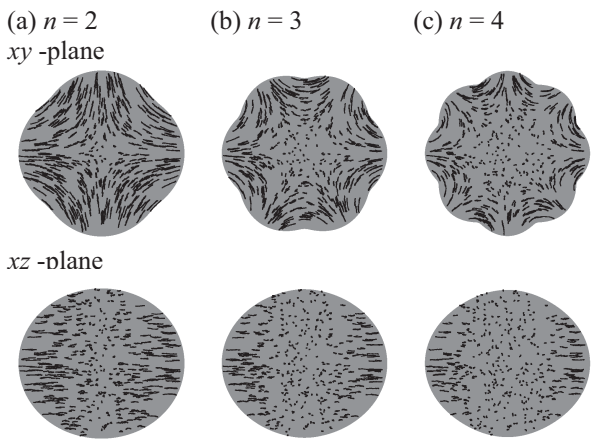}
	\caption{Trajectories of 400 tracer particles for one cycle (from $t=0$ to $T$) on $xy$- (top) and $xz$- (bottom) planes. The particles move back and forth with the reciprocal flow. However, they do not follow the streamlines due to the thermal noise. As the initial condition, the particles are randomly located on each plane. The oscillation amplitude $\varepsilon$ is set as 0.1. The gray regions indicate the area of the droplet, where the time-dependent shapes are all superimposed.}
	\label{fig_imposed}
\end{figure}

In Fig.~\ref{fig_imposed}, the trajectories of the tracer particles on the $xy$- and $xz$- planes during a period are shown for $n = 2, 3$, and $4$ to illustrate the particle motion and flow field. We set the oscillation amplitude as $\varepsilon = 0.1$. The particles are confined in the droplet by recalculating the noise till the next particle position is located inside the droplet.

We prepared the particles at a distance $\rho$ from the center on the $xy$-plane with 100 different initial angles, i.e., $\varphi = 2 \pi k / 100$ for $k = 0, \cdots, 99$.
To obtain the diffusion tensor in Eq.~\eqref{D_M}, we calculate each component of the second moment $M^{(2)} (\bm{r})$ in Eq.~\eqref{def_KM2}, by averaging the displacements of $10^7$ particles for each initial location.
Then, it was translated into the diffusion tensor in the cylindrical coordinates by using the relations in Eqs.~\eqref{def_drr} to \eqref{def_dphiz} in Appendix~\ref{exp_D}.

\begin{figure}
	\centering
	\includegraphics{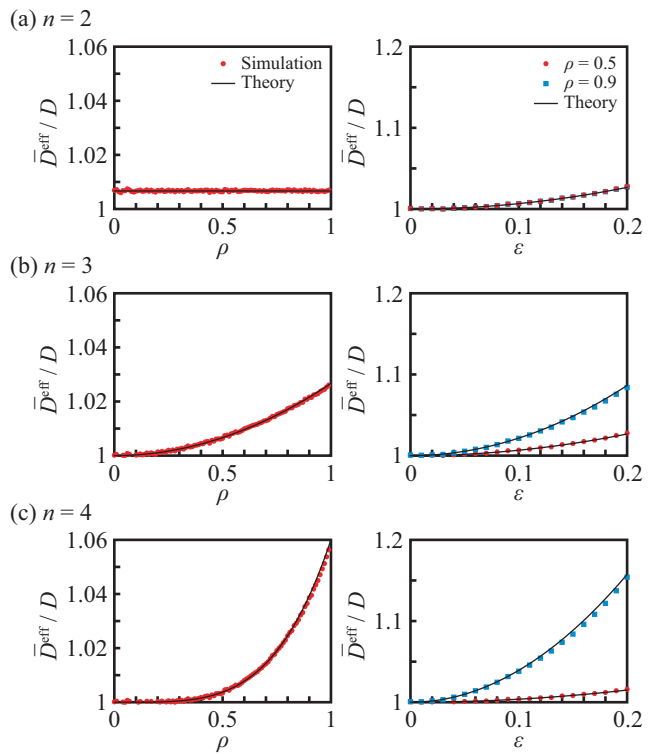}
	\caption{Numerical results of the normalized effective diffusion coefficient $\bar{D}^{\mathrm{eff}}/D$ dependent on the initial distance $\rho$ from the center (left) and the oscillation amplitude $\varepsilon$ (right) on the $xy$ plane for $n = 2, 3$, and $4$. The effective diffusion coefficient was estimated from the mean square displacement of the particles. With respect to the left panel, $\varepsilon$ is fixed at $\varepsilon = 0.1$, and for the right panel, the initial radius is fixed at $\rho = 0.5$ [red (gray)] and $\rho = 0.9$ [blue (light gray)]. The black thin curves show the theoretical prediction in Eq.~\eqref{D_eff}.}
	\label{fig_rad_epsilon}
\end{figure}

Based on the obtained diffusion tensor, the effective diffusion coefficient in Eq.~\eqref{eff_diff} was calculated.
The dependence of the normalized diffusion coefficient $\bar{D}^{\mathrm{eff}}/D$ on the initial distance from the center $\rho$ is plotted in the left panel of Fig.~\ref{fig_rad_epsilon}.
The dependence of the normalized diffusion coefficient $\bar{D}^{\mathrm{eff}}/D$ on the oscillation amplitude $\varepsilon$ was also calculated for the initial distance from the center $\rho = 0.5$ and $0.9$, which is shown in the right panel of Fig.~\ref{fig_rad_epsilon}.
The effective diffusion increases as $\varepsilon$ increases for $n = 2, 3$, and $4$.
In Fig.~\ref{fig_rad_epsilon}, the theoretical result in Eq.~\eqref{D_eff} is illustrated by black curves.
The effective diffusion coefficient obtained by the numerical calculation is in good agreement with the theoretical prediction.

\begin{figure}
	\centering
	\includegraphics{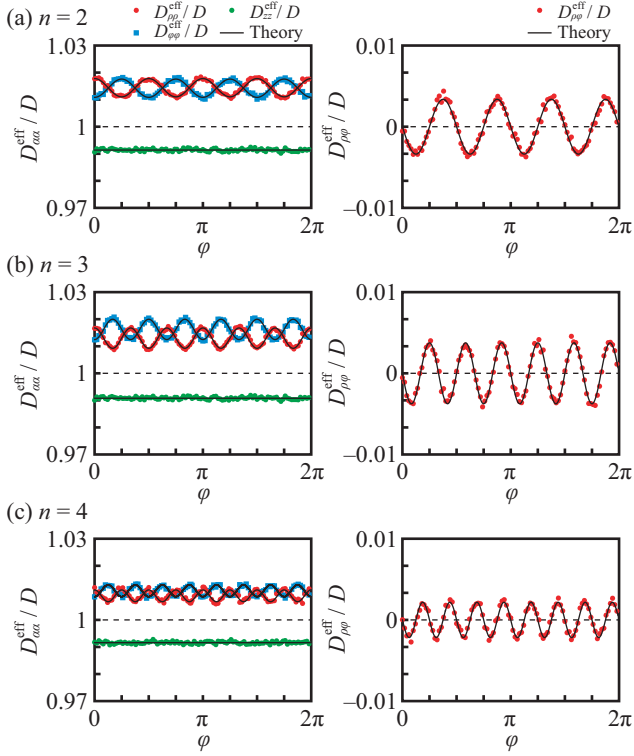}
	\caption{Numerical results on the components of the normalized effective diffusion coefficient dependent on $\varphi$ for $n = 2, 3$, and $4$. $D^{\mathrm{eff}}_{\rho\rho}/D$ [red (gray)], $D^{\mathrm{eff}}_{\varphi \varphi}/D$ [blue (light gray)], $D^{\mathrm{eff}}_{zz}/D$ [green (dark gray)] are shown in the left panel, while $D^{\mathrm{eff}}_{\rho\varphi}/D$ is plotted on the right panel. The initial distance from the center is $\rho = 0.5$, and $\varepsilon = 0.1$.
	The black thin curves show the theoretical predictions in Eqs.~\eqref{drr} to \eqref{drphi}.}
	\label{fig_angle_dep}
\end{figure}

To obtain the $\varphi$ dependence of the components of the normalized effective diffusion tensor, $D^{\mathrm{eff}}_{\rho\rho}/D$, $D^{\mathrm{eff}}_{\varphi \varphi}/D$, $D^{\mathrm{eff}}_{zz}/D$, and $D^{\mathrm{eff}}_{\rho\varphi}/D$, the numerical calculations were performed from the initial condition $\rho = 0.5$.
The averaging was performed for $10^7$ particles from each initial position. The results are also shown in Fig.~\ref{fig_angle_dep} for $n = 2, 3$, and $4$. 
The figure shows that $D^{\mathrm{eff}}_{\rho\rho}$ and $D^{\mathrm{eff}}_{\varphi \varphi}$ are greater than the thermal equilibrium diffusion coefficient $D$, and they depend on $\varphi$ with a wave number of $2n$.
On the while, $D^{\mathrm{eff}}_{zz}$ is smaller than $D$, and is independent of $\varphi$. The normalized effective cross diffusion coefficient $D^{\mathrm{eff}}_{\rho\varphi}$ is also periodic with a wavenumber of $2n$.
In Fig.~\ref{fig_angle_dep}, the theoretical results in Eqs.~\eqref{drr} to \eqref{drphi} are also illustrated by black curves.
The components of the effective diffusion tensor obtained via numerical calculation are in good agreement with the theoretical prediction.

\section{Discussion}

In our model, the viscous term is omitted from Eq.~\eqref{NS_nd}, and Eq.~\eqref{NS_mod-} is adopted.
The nonlinear term, $(\bm{v} \cdot \nabla) \bm{v}$, is assumed to be smaller than the other terms in Eq.~\eqref{NS_mod-} in the theoretical calculation.
In the experiments in Ref.~\cite{Watanabe}, the characteristic time scale was 6.5 ms, which is the oscillation period of the droplet deformation.
The characteristic velocity of the flow is estimated by the amplitude over the oscillation period: 0.2 mm / 6.5 ms $\sim$ 0.03 m/s.
The density, viscosity, and surface tension of water are $10^3$ kg/$\mathrm{m}^3$, $10^{-3}$ Pa~s, $7 \times 10^{-2}$ N/m, respectively.
Based on these values, $\varepsilon$, $\sigma$, and $1/\mathrm{Re}$ are calculated as follows:
\begin{align}
\varepsilon &\sim 2 \times 10^{-1}, \\
\sigma &\sim 3, \\
\frac{1}{\mathrm{Re}} &\sim 3 \times 10^{-2}.
\end{align}
Here, the symbol ``$\sim$'' denotes that the terms connected with the symbol are in the same order.

The order of the terms in Eq.~\eqref{NS_nd} are estimated as follows:
We recall that $|\bm{v}|$ is the order of $\varepsilon$, and assume that the order of $\left | \nabla p \right |$ is the order of the gradient of the Laplace pressure, $|\sigma \nabla H|$.
The order of $\nabla H$ is evaluated by the difference between the maximum and minimum values of $H$ as follows:
\begin{align}
|\nabla H| \sim \left ( \frac{1}{R+\Delta R}-\frac{1}{R-\Delta R} \right ) \sim \frac{2\varepsilon}{R},
\end{align}
where we used the form with dimension.
Then, we have
\begin{align}
\left | \frac{\partial \bm{v}}{\partial t} \right | &\sim \varepsilon, \\
\left | \bm{v} \cdot \nabla \bm{v} \right | &\sim \varepsilon^2, \\
\left | \nabla p \right | &\sim \sigma \varepsilon, \\
\left | \frac{1}{\mathrm{Re}} \nabla^2 \bm{v} \right | &\sim \frac{\varepsilon}{\mathrm{Re}},
\end{align}
that is to say,
\begin{align}
\left | \frac{1}{\mathrm{Re}} \nabla^2 \bm{v} \right | < \left | \bm{v} \cdot \nabla \bm{v} \right | < \left | \frac{\partial \bm{v}}{\partial t} \right | 
 \sim \left | \nabla p \right |.
\end{align}
Thus, the adopted assumptions are valid.

It should be noted that our approach is not applicable for $n=5$ and $n=10$, because $c_{nk}$, $g_{nk}$, and $h_{nk}$ in Eqs.~\eqref{drr}, \eqref{dphiphi}, and \eqref{dzz} diverge for $(n,k) =(5,4)$ and $(10,8)$.
This is because of the resonance, wherein the second harmonic of the considered mode is the same as the other characteristic mode. Actually, $2\omega_5 = \omega_8$ and $2\omega_{10} = \omega_{16}$. We confirmed that pairs of positive integers $(n, k)$ that satisfy $2\omega_n = \omega_{2k}$ are $(5,4)$ and $(10,8)$ in the range of $2 \le n \le 100$.

In actual systems, the aforementioned divergence of oscillation amplitude will not occur due to the following reasons:
First, if the amplitude of the deformation oscillation becomes larger, then perturbative treatment is not effective. Thus, the present approach cannot be adopted.
Second, energy dissipation plays an essential role in the case of larger deformations.
As we discussed, the energy dissipation is relatively smaller than the other effects, such as inertia or pressure.
In our model, we neglected the energy dissipation. However, it cannot be neglected when the amplitude becomes large.
It is known that when the energy dissipation and injection balance, the amplitude of the oscillation mode becomes finite.
Thus, our model can be adopted if we extend our model to include energy dissipation and injection.
Third, the droplet deforms to an ellipsoidal shape due to the gravity effect and anisotropy of the acoustic field in experiments.
In such cases, the characteristic frequency should be shifted~\cite{Tsamopoulos}, and thus the resonance will not occur.

\section{Summary}

The dynamics of mixing, which involve flow and diffusion processes, are usually described by the convection-diffusion equation~\eqref{FP}.
A method to investigate the cooperative dynamics of flow and diffusion involves tracking the time evolution of the convection-diffusion equation. However, it is difficult to determine the position-dependent diffusion coefficient from the dynamics of the concentration field.
In the present study, we derive the discrete time evolution equation per the period of the reciprocal flow, as shown in Eq.~\eqref{dFP}.
Since the reciprocal flow does not induce net convection in a period, the convection term disappears, as shown in Eq.~\eqref{dfp.eq}.
Instead, we can directly observe the effective diffusion coefficient per the period. Thus, we succeeded in showing that the reciprocal flow, which cannot agitate a fluid, affects diffusion.

The Brownian motion causes the transition of a tracer particle between fluid elements.
The probability of a single transition is equal to that of an inverted process, which results in normal diffusion.
When a flow is stimulated, the fluid elements deform in time.
Thus, a sequent transition process due to thermal fluctuation becomes irreversible and realizes an anisotropic Brownian motion.
The diffusion enhancement is induced by such anisotropic Brownian motion.

The representation of the effective diffusion tensor indicates that the diffusion process can be affected by any reciprocal flow.
In the present system with a levitated droplet, the diffusion is enhanced by the reciprocal flow. However, it is not clear whether any reciprocal flow always enhances the diffusion.
Hence, this can be explored in a future study.

\begin{acknowledgments}
The authors acknowledge Ayumu Watanabe and Suguru Komaya for shearing the latest experimental results.
We also acknowledge Sakurako Tanida for helpful discussion.
This work was supported by JSPS KAKENHI Grant Nos.~JP19J00365, JP19K0376, and JP20K14370.
This work was also supported by JSPS and PAN under the Japan-Poland Research Cooperative Program ``Spatio-temporal patterns of elements driven by self-generated, geometrically constrained flows'' and ``Complex spatio-temporal structures emerging from interacting self-propelled particles'' (JPJSBP120204602), and the Cooperative Research Program of ``Network Joint Research Center for Materials and Devices'' (Nos.~20191030, 20194006, 20201023, and 20204004).
\end{acknowledgments}

\appendix

\section{Explicit form of mean curvature $H$\label{mc}}

If the shape of the droplet is described as:
\begin{align}
\bm{r}_b &= R \left[ 1 + \epsilon s(\theta, \varphi, t)\right] \bm{e}_r, \end{align}
then the mean curvature $H$ up to the order of $\epsilon^2$ is calculated as follows: 
\begin{align}
&H = -\frac{1}{R} + \frac{\epsilon}{R} \left[s + \frac{1}{2} \frac{\partial^2 s}{\partial \theta^2} + \frac{\cos\theta}{2\sin\theta} \frac{\partial s}{\partial \theta} + \frac{1}{2\sin^2\theta}\frac{\partial^2 s}{\partial\varphi^2} \right] \nonumber \\
&- \frac{\epsilon^2}{R} s \left[ s + \frac{\partial^2 s}{\partial \theta^2} + \frac{\cos\theta}{\sin\theta} \frac{\partial s}{\partial \theta} + \frac{1}{\sin^2\theta}\frac{\partial^2 s}{\partial\varphi^2} \right] + \mathcal{O} (\epsilon^3),
\end{align}
where $\epsilon$ is a small parameter.

\section{Equation and boundary conditions with respect to $\varepsilon$\label{eq_bc}}

We show the equations and boundary conditions with respect to a small parameter $\varepsilon$.
By substituting Eqs.~\eqref{f_expand}, \eqref{Phi_expand}, and \eqref{p_expand} to Eq.~\eqref{NS_mod} and the boundary conditions \eqref{LaplacePressure_mod} and \eqref{bc_shape-}, we obtain the equations for the order of $\varepsilon$ as follows:
\begin{align}
\frac{\partial \Phi^{(1)}}{\partial t} = - p^{(1)}, \label{NS1}
\end{align}
\begin{widetext}
\begin{align}
\left. p^{(1)} \right|_{r=1} 
=& -\sigma \left[f^{(1)}(\theta, \varphi, t) + \frac{1}{2} \frac{\partial^2 f^{(1)}}{\partial \theta^2} + \frac{\cos\theta}{2\sin\theta} \frac{\partial f^{(1)}}{\partial \theta} + \frac{1}{2\sin^2\theta}\frac{\partial^2f^{(1)}}{\partial\varphi^2} \right],
\end{align}
\begin{align}
\left. \nabla \Phi^{(1)} \right|_{r = 1} \cdot \bm{e}_r = \frac{\partial f^{(1)}}{\partial t}. \label{bc1}
\end{align}

The equations of the order of $\varepsilon^2$ are as follows:
\begin{align}
\frac{\partial \Phi^{(2)}}{\partial t} + \frac{1}{2} \left| \nabla \Phi^{(1)} \right|^2 = - p^{(2)}, \label{NS2}
\end{align}
\begin{align}
\left. p^{(2)} \right|_{r=1} + \left. \frac{\partial p^{(1)}}{\partial r} \right|_{r = 1} f^{(1)}(\theta, \varphi, t) =& \sigma f^{(1)}(\theta,\varphi, t) \left[ f^{(1)}(\theta, \varphi, t) + \frac{\partial^2 f^{(1)}}{\partial \theta^2} + \frac{\cos\theta}{\sin\theta} \frac{\partial f^{(1)}}{\partial \theta} + \frac{1}{\sin^2\theta}\frac{\partial^2f^{(1)}}{\partial\varphi^2} \right] \nonumber \\
&- \sigma \left[f^{(2)}(\theta, \varphi, t) + \frac{1}{2} \frac{\partial^2 f^{(2)}}{\partial \theta^2} + \frac{\cos\theta}{2\sin\theta} \frac{\partial f^{(2)}}{\partial \theta} + \frac{1}{2\sin^2\theta}\frac{\partial^2f^{(2)}}{\partial\varphi^2} \right], \label{bc2}
\end{align}
\begin{align}
\left. \nabla \Phi^{(2)} \right|_{r = 1} \cdot \bm{e}_r + \left. \left( \frac{\partial}{\partial r} ( \nabla \Phi^{(1)} ) \right) \right|_{r = 1} \cdot \bm{e}_r f^{(1)}(\theta,\varphi, t)
 - \left. \nabla \Phi^{(1)} \right|_{r = 1} \cdot \left( \frac{\partial f^{(1)}}{\partial \theta} \bm{e}_\theta + \frac{1}{\sin \theta} \frac{\partial f^{(1)}}{\partial \varphi} \bm{e}_\varphi \right) = \frac{\partial f^{(2)}}{\partial t}. \label{phi2}
\end{align}
It should be noted that $p^{(1)}$ and $p^{(2)}$ are not unique.
Hence, any constant value can be added.
\end{widetext}

\section{Explicit forms of the coefficients in Eqs.~\eqref{r2exp} and \eqref{Phi2exp}\label{sol}}

The explicit forms of the coefficients in Eqs.~\eqref{r2exp} and \eqref{Phi2exp} are shown as follows:
\begin{align}
\hat{f}^{(2,n)}_{2n,2n} = \frac{n^3 - 2n^2 -5n + 4}{8 (n^2 + 1)},
\end{align}
\begin{align}
\bar{f}^{(2,n)}_{2n,2n} = \frac{n^3 + 3n^2 -2}{4(2n -1)(2n + 2)},
\end{align}
\begin{align}
\hat{f}^{(2,n)}_{2k,0} = \frac{ Y(n,k)}{\left[(2n-1)!!\right]^2},
\end{align}
\begin{align}
\bar{f}^{(2,n)}_{2k,0} = \frac{Z(n,k)}{[(2n-1)!!]^2},
\end{align}
\begin{align}
\hat{\Phi}^{(2,n)}_{2n,2n} = -\frac{3 \omega_n (n+1)(n-1)^2 }{8n (n^2 + 1)}, \label{Phi2n2n}
\end{align}
\begin{align}
\hat{\Phi}^{(2,n)}_{2k,0} = -\frac{\omega_n \Xi(n,k)}{\left[(2n-1)!!\right]^2}. \label{Phi2n2k0}
\end{align}
Here, we set
\begin{widetext}
\begin{align}
Y(n,k) =& \frac{(-1)^k (4k + 1) \left[(2n)!\right]^2 (2k -1)!! \left[(4n^2 - 4k^2n + 4kn^2 - 2k^3 -k^2 -kn)(n-1)(n+2) + 4kn(n^2+n-1) \right]}{ 2^{n+4} n (k - k^2 - 2 k^3 - 2 n + n^2 + n^3)(2n + 2k + 1)!! (n - k)! k !},
\end{align}
\begin{align}
Z(n,k) =& \frac{(-1)^k (4k + 1) \left[(2n)!\right]^2 (2k -1)!! \left[k (2 k^2 + k -n)(n-1)(n+2) + 4kn(n^2+n-1) \right] }{2^{n+4} k (2k -1) n(2n + 2k + 1)!! (n - k)! (k+1)!}.
\end{align}
\begin{align}
\Xi(n,k) =& \frac{(-1)^{k-1} (4k + 1) \left[(2n)!\right]^2 (2k -1)!!}{2^{n+4} n (k - k^2 - 2 k^3 - 2 n + n^2 + n^3)(2n + 2k + 1)!! (n - k)! k ! } \nonumber \\
& \times \left[8 k^3 + 8 k^4 + k (n + 1) (n - 4) - n (n - 1) (n + 2) (4 n + 5) + 2 k^2 (n^2 - 3 n - 3) \right],
\end{align}
\end{widetext}

\noindent \\

\section{Explicit forms of the diffusion coefficient\label{exp_D}}

The diffusion tensor in the cylindrical coordinates is obtained as:
\begin{align}
D^{\mathrm{eff}}_{\rho\rho} =& D_{xx}^{\mathrm{eff}} \cos^2 \varphi + D_{yy}^{\mathrm{eff}} \sin^2 \varphi + 2 D_{xy}^{\mathrm{eff}} \sin \varphi \cos \varphi, \label{def_drr} \\
D^{\mathrm{eff}}_{\varphi \varphi} =& D_{xx}^{\mathrm{eff}} \sin^2 \varphi + D_{yy}^{\mathrm{eff}} \cos^2 \varphi - 2 D_{xy}^{\mathrm{eff}} \sin \varphi \cos \varphi, \label{def_dphiphi} \\
D^\mathrm{eff}_{\rho \varphi} =& 2 \left( D_{yy}^{\mathrm{eff}} - D_{xx}^{\mathrm{eff}} \right) \sin \varphi \cos \varphi \nonumber \\
& + 2 D_{xy}^{\mathrm{eff}} \left(\cos^2 \varphi - \sin^2 \varphi\right),\label{def_drphi} \\
D^\mathrm{eff}_{\rho z} =& \left ( D_{xz}^{\mathrm{eff}} \cos \varphi + D_{yz}^{\mathrm{eff}} \sin \varphi \right ), \label{def_drz} \\
D^\mathrm{eff}_{\varphi z} =& \left ( -D_{xz}^{\mathrm{eff}} \sin \varphi + D_{yz}^{\mathrm{eff}} \cos \varphi \right ). \label{def_dphiz}
\end{align}

The coefficients in Eqs.~\eqref{drr} to \eqref{drphi} are explicitly described as follows:
\begin{align}
b_n &= \frac{n(n-1)}{2}, \\
c_{nk} &= \frac{(-1)^k (2k-1) (2k-1)! \Xi(n,k)}{2^{2k-2}[(k-1)!]^2 [(2n-1)!!]^2 }, \\
d_{n} &= \frac{3(n+1)(n-1)^2(2n-1)}{4(n^2+1)}, \\
g_{nk} &= \frac{(-1)^k (2k-1)! \Xi(n,k)}{2^{2k-2}[(k-1)!]^2 [(2n-1)!!]^2},\\
h_{nk} &= \frac{(-1)^{k-1} (2 k)! \Xi(n,k)}{2^{2k-2} [(k-1)!]^2 [(2n-1)!!]^2}.
\end{align}

The explicit forms of the effective diffusion coefficient $\bar{D}^{\mathrm{eff}}$ and the components of the effective diffusion tensor in the $xy$-plane for modes $n = 2, 3$, and $4$ are shown as follows:
For $n = 2$, we have
\begin{align}
\frac{\bar{D}^{\mathrm{eff}}}{D} = 1 + \frac{2}{3} \varepsilon^2,
\end{align}
\begin{align}
\frac{D^{\mathrm{eff}}_{\rho\rho}}{D} = 1 + \varepsilon^2 \left[ \frac{89}{63} + \frac{81}{700} \rho^2 + \frac{27}{20} \rho^2 \cos (4\varphi) \right],
\end{align}
\begin{align}
\frac{D^{\mathrm{eff}}_{\varphi \varphi}}{D} = 1+ \varepsilon^2 \left[ \frac{89}{63} + \frac{27}{700} \rho^2 - \frac{27}{20} \rho^2 \cos(4\varphi) \right],
\end{align}
\begin{align}
D^{\mathrm{eff}}_{zz}= 1 + \varepsilon^2 \left[ -\frac{52}{63} - \frac{27}{175} \rho^2 \right],
\end{align}
\begin{align}
D^{\mathrm{eff}}_{\rho\varphi}= \varepsilon^2 \left[ -\frac{27}{20} \rho^2 \sin (4\varphi) \right].
\end{align}
For $n = 3$, we have
\begin{align}
\frac{\bar{D}^{\mathrm{eff}}}{D} = 1 + \frac{8}{3} \varepsilon^2 \rho^2,
\end{align}
\begin{align}
\frac{D_{\rho\rho}^{\mathrm{eff}}}{D} =& 1 + \varepsilon^2 \left[ \frac{2}{7} + \frac{177}{44} \rho^2 + \frac{10}{77} \rho^4 + 6 \rho^4 \cos (6\varphi)\right],
\end{align}
\begin{align}
\frac{D_{\varphi \varphi}^{\mathrm{eff}}}{D} =& 1 + \varepsilon^2 \left[\frac{2}{7} + \frac{235}{44} \rho^2 + \frac{2}{77} \rho^4 - 6 \rho^4 \cos(6\varphi) \right],
\end{align}
\begin{align}
\frac{D_{zz}^{\mathrm{eff}}}{D} = 1 + \varepsilon^2 \left[ -\frac{4}{7} - \frac{15}{11} \rho^2 - \frac{12}{77} \rho^4 \right],
\end{align}
\begin{align}
\frac{D_{\rho\varphi}^{\mathrm{eff}}}{D} = \varepsilon^2 \left[ -6 \rho^4 \sin (6\varphi) \right].
\end{align}
For $n = 4$, we have
\begin{align}
\frac{\bar{D}^{\mathrm{eff}}}{D}= 1 + 6 \varepsilon^2 \rho^4,
\end{align}
\begin{align} \frac{D_{\rho\rho}^{\mathrm{eff}}}{D}=& 1 + \varepsilon^2 \left[ \frac{664}{2695} + \frac{3936}{5005} \rho^2 + \frac{157}{22} \rho^4 \right. \nonumber \\ & \left. + \frac{735}{9724} \rho^6 + \frac{945}{68} \rho^6 \cos (8\varphi)\right],
\end{align}
\begin{align} \frac{D_{\varphi \varphi}^{\mathrm{eff}}}{D} =& 1 + \varepsilon^2 \left[ \frac{664}{2695} + \frac{1312}{5005} \rho^2 + \frac{269}{22} \rho^4 \right. \nonumber \\ & \left. + \frac{105}{9724} \rho^6 - \frac{945}{68} \rho^6 \cos(8\varphi)\right],
\end{align}
\begin{align} \frac{D_{zz}^{\mathrm{eff}}}{D}=& 1 + \varepsilon^2 \left[ -\frac{1328}{2695} - \frac{5248}{5005} \rho^2 - \frac{15}{11} \rho^4 - \frac{210}{2431} \rho^6 \right],
\end{align}
\begin{align} \frac{D_{\rho\varphi}^{\mathrm{eff}}}{D} = \varepsilon^2 \left[-\frac{945}{68} \rho^6 \sin (8\varphi)\right].
\end{align}


\begin{thebibliography}{99}

\bibitem{Ishikawa}
T.~Ishikawa, P-F.~Paradis, J.~T.~Okada, and Y.~Watanabe, \textit{Meas. Sci. Technol.} \textbf{23}, 025305 (2012).

\bibitem{Rhim} W.~Rhim and S.~K.~Chung, \textit{Methods} \textbf{1}, 118 (1990). 

\bibitem{Brillo} J.~Brillo, A.~I.~Pommrich, and A.~Meyer, \textit{Phys. Rev. Lett.} \textbf{107}, 165902 (2011).

\bibitem{Liua}
Y.~Liu, D.~Zhu, D.~M.~Strayer, and U.~Israelsson, \textit{Adv. Space Res.} \textbf{45}, 208 (2010).

\bibitem{Hill} R.~J.~A.~Hill and L.~Eaves, \textit{Phys. Rev. Lett.} \textbf{101}, 234501 (2008).

\bibitem{Mestel} A.~J.~Mestel, \textit{J. Fluid. Mech.} \textbf{117}, 27 (1982).

\bibitem{Marzo} A.~Marzo, S.~A.~Seah, B.~W.~Drinkwater, D.~R.~Sahoo, B.~Long, and S.~Subramanian, \textit{Nat. Comm.} \textbf{6}, 8661 (2015).

\bibitem{Hirayama} R.~Hirayama, D.~M.~Plasencia, N.~Masuda, and S.~Subramanian, \textit{Nature} \textbf{575}, 320 (2019).

\bibitem{Morris} R.~H.~Morris, E.~R.~Dye, P.~Docker, and M.~I.~Newton, \textit{Phys. Fluids} \textbf{31}, 101301 (2019).

\bibitem{Sasaki} Y.~Sasaki, K.~Kobayashi, K.~Hasegawa, A.~Kaneko, and Y.~Abe, \textit{Phys. Fluids} \textbf{31}, 102109 (2019).

\bibitem{OhsakaPRL} K.~Ohsaka and E.~H.~Trinh, \textit{Phys. Rev. Lett.} \textbf{84}, 1700 (2000).
\bibitem{Feng} J.~Q.~Feng and K.~V.~Beard, \textit{J. Atmos. Sci.} \textbf{48}, 1856 (1991).

\bibitem{Price} C.~J.~Price, T.~D.~Donnelly, S.~Giltrap, N.~H.~Stuart, S.~Parker, S.~Patankar, H.~F.~Lowe, D.~Drew, E.~T.~Gumbrell, and R.~A.~Smith. \textit{Rev. Sci. Instrum.} \textbf{86}, 033502 (2015).

\bibitem{DynamicsDroplet} A.~Frohn and N.~Roth, \textit{Dynamics of Droplets} (Springer, Berlin, 2000).

\bibitem{Watanabe2008} T.~Watanabe, \textit{Phys. Lett. A} \textbf{372}, 482 (2008).

\bibitem{KitahataPRE} H.~Kitahata, R.~Tanaka, Y.~Koyano, S.~Matsumoto, K.~Nishinari, T.~Watanabe, K.~Hasegawa, T.~Kanagawa, A.~Kaneko, and Y.~Abe, \textit{Phys. Rev. E} \textbf{92}, 062904 (2015).

\bibitem{Watanabe2010} T.~Watanabe, \textit{Int. J. Geol.} \textbf{4}, 5 (2010).

\bibitem{Watanabe} A.~Watanabe, K.~Hasegawa, and Y.~Abe, \textit{Sci. Rep.} \textbf{8}, 10221 (2018).

\bibitem{Hasegawa} K.~Hasegawa, A.~Watanabe, A.~Kaneko, and Y.~Abe, \textit{Phys. Fluids} \textbf{31}, 112101 (2019).

\bibitem{Purcell} E.~M.~Purcell, \textit{Am. J. Phys.} \textbf{45}, 3 (1977).

\bibitem{Stokes1847} G.~G.~Stokes, \textit{Trans. Camb. Phil. Soc.} \textbf{8}, 441-455 (1847).

\bibitem{deGennes} P-G. de Gennes, F.~Brochard-Wyart, D.~Quere, \textit{Capillarity and Wetting Phenomena: Drops, Bubbles, Pearls, Waves} (Springer, New York, 2004).

\bibitem{Rayleigh} L.~Rayleigh, \textit{Proc. Royal Soc. Lond.} \textbf{29}, 71 (1879).

\bibitem{Risken} H.~Risken and T.~Frank, \textit{The Fokker-Planck Equation: Methods of Solution and Applications} (Springer, Berlin, 1996).

\bibitem{Synergetics2} A.~S.~Mikhailov and A.~Y.~Loskutov, \textit{Foundations of Synergetics II: Complex Patterns} (Springer-Verlag, Berlin, 1991).

\bibitem{Lamb} H.~Lamb, \textit{Hydrodynamics} (Cambridge University Press, Cambridge, England, 1895).

\bibitem{Landau}L.~D.~Landau and E.~M.~Lifshitz, \textit{Fluid Mechanics} (Pergamon Press, Oxford, 1959).

\bibitem{Numerical_Recipes} W.~H.~Press, S.~A.~Teukolsky, W.~T.~Vetterling, and B.~P.~Flannery, \textit{Numerical Recipes: The Art of Scientific Computing} (Cambridge University Press, Cambridge, England, 2007).

\bibitem{Tsamopoulos} J.~A.~Tsamopoulos and R.~A.~Brown, \textit{J. Fluid. Mech.} \textbf{127}, 519 (1983).

\end{thebibliography}
\end{document}